\documentclass[12pt]{article}

\newcommand{\be}{\begin{equation}}
\newcommand{\ee}{\end{equation}}
\newcommand{\bea}{\begin{eqnarray}}
\newcommand{\eea}{\end{eqnarray}}

\newcommand{\nn}{\nonumber\\}

\newcommand{\bml}{\begin{multline}}
\newcommand{\eml}{\end{multline}}

\begin{document}

\begin{center}

{\Large {\bf Non-perturbative time-dependent String Backgrounds and Axion-induced Optical Activity}}

\vspace{1.5cm}

Jean Alexandre\footnote{jean.alexandre@kcl.ac.uk}, Nick E. Mavromatos\footnote{nikolaos.mavromatos@kcl.ac.uk} and
Dylan Tanner\footnote{dylan.tanner@kcl.ac.uk} \\

\vspace{0.5cm}

{\it Department of Physics,\\ King's College London, \\ London WC2R 2LS, U.K.}

\vspace{2cm}

{\bf Abstract}

\end{center}

Conformal invariance for bosonic strings in time-dependent backgrounds of graviton, dilaton and Kalb-Ramond field is
obtained by imposing Weyl-beta functions to be homogeneous in time, to all orders in $\alpha^{'}$. This construction
is possible in any target space
dimension, as a result of the non-trivial background configurations.
The electromagnetic effects of the antisymmetric tensor field, when coupled to an Abelian (electromagnetic) gauge field,
are discussed in the framework of a
four-dimensional Minkowski Universe, for concreteness. Non-trivial optical activity is
demonstrated, which constitutes a way of detecting string-inspired axion/dilaton effects in such models.

\vspace{1cm}

{\it Introduction.} The target-space dynamics of strings, in a first-quantized framework, is based on equations of motion, involving backgrounds of the massless degrees of freedom of the string, which are generated by the vanishing of the Weyl-invariance beta functions
$\beta^i$ of the world sheet non-linear sigma model. The usual approach consists of seeking for
perturbative solutions of the vanishing of these beta functions, truncated at most to second order in $\alpha^{'}$. Although such an approach may suffice for a discussion of low-energy (field-theoretic) effects of strings, non-perturbative (to all orders in $\alpha '$) re-summation schemes may be essential, in particular when discussing effects at the early Universe, or in cases where the string mass scale is sufficiently low, e.g. of the order of a few TeV, as is the case in some of the modern approaches to string theories.

To this end, an  approach to obtaining re-summed in $\alpha '$ Weyl anomaly coefficients was proposed
in \cite{AEM1}, in the case of a time-dependent configuration of the bosonic string in metric and dilaton backgrounds. This approach, which is based on a novel functional method on the world-sheet of the string,
leads to homogeneous $\beta^i$: for each of these, all orders in $\alpha^{'}$ are the same power law in time. This suffices to secure conformal invariance as follows: From two-loop (in $\sigma$-model) and higher, the expressions for the beta functions are not unique, but can be modified by a set of
field redefinitions, i.e. a string reparametrization \cite{metsaev} which does not affect the physical
predictions of the theory. As a consequence, if all orders in $\alpha^{'}$ of $\beta^i$ are homogeneous in time
(same power law $(X^0)^{a_i}$), each beta function can be written in the form
\be
\beta^i=\sum_{n=0}^\infty\xi_n^i(X^0)^{a_i}(\alpha^{'})^n=A^i(X^0)^{a_i},
\ee
where the constant $A^i$ depends on
the string parametrization, and can be set to zero by choosing specific field redefinitions. This cancellation was
explicitly shown at two loops in \cite{AEM1}, and arguments have been given that it can be performed to all orders. It was found in \cite{AEM1} that the effect of this string reparametrization is to rescale the metric by a constant, and add another constant to the dilaton, and therefore does not change the time dependence of the configuration. We note that a string reparametrization, generated by field redefinitions, corresponds to changing the renormalization scheme for the calculation of the renormalized energy momentum tensor of the world sheet sigma model.

The target space corresponding to the configuration found in \cite{AEM1} is a power law expanding
Universe, whose dimension $D$ is not restricted by any constraint, since the abovementioned non-perturbative
conformal invariance does not involve the vanishing of the tree-level conformal anomaly $(D-26)/6$. The power law
scale factor becomes a constant, leading to a Minkowski Universe, for a specific choice of the dilaton
amplitude, depending on $D$.

An extension of the work in \cite{AEM1}, involving the antisymmetric tensor was made
in \cite{AMT1} and is based on a similar non-perturbative conformal invariance, which is different from the traditional approach \cite{copeland}. We study here the effect of the antisymmetric tensor in a four-dimensional Minkowski space time (ref.~\cite{AMT1} deals with a linearly expanding space time), in order to take into account more realistically the observed anisotropy in the propagation of electromagnetic waves on cosmological distances \cite{nodland}.

{\it Conformal properties.} To first order in $\alpha^{'}$, the beta functions for the bosonic world-sheet
$\sigma$-model theory in graviton, antisymmetric tensor and dilaton backgrounds are \cite{metsaev}:
\bea\label{weylconditions}
\beta_{\mu\nu}^{g(1)}&=&R_{\mu\nu}+2\nabla_\mu\nabla_\nu\phi-\frac{1}{4}H_{\mu\rho\sigma}H_\nu^{~\rho\sigma}\\
\beta_{\mu\nu}^{B(1)}&=&-\frac{1}{2}\nabla^\rho H_{\rho\mu\nu}+H_{\rho\mu\nu}\partial^\rho\phi\nn
\beta^{\phi(1)}&=&\frac{D-26}{6\alpha^{'}}-\frac{1}{2}\nabla^2\phi+\partial^\rho\phi\partial_\rho\phi
-\frac{1}{24}H_{\mu\nu\rho}H^{\mu\nu\rho}\nonumber
\eea
Exploiting the consistency of our conformal-invariant solutions to any target-space dimension~\cite{AEM1},
 we will restrict our study to four space-time dimensions, where we do not need any compactification procedure
or confinement on a three-spatial-dimensional brane. The most general time-dependent three-form $H$ is then
\be\label{axion}
H_{\rho\mu\nu}(X^0)=(X^0)^m\varepsilon_{\rho\mu\nu\sigma}(X^0)\partial^\sigma \left( \sum_{i=1}^3h_iX^i\right) ,
\ee
where $h_i$ are constants and $m$ is a power to be fixed later on.
This expression defines a four-dimensional pseudoscalar axion field $b = \sum_{i=1}^3h_iX^i $, in the notation of
\cite{ABEN}.

The metric is chosen as
\be
g_{\mu\nu}=\mbox{diag}\left( \kappa(X^0),\tau_1(X^0),\tau_1(X^0),\tau_3(X^0)\right),
\ee
and since
$\varepsilon_{0123}=\sqrt{\mid g(X^0)\mid}$, we have
\be
H_{0ij}(X^0)=(X^0)^m\sqrt{|\kappa(X^0)\tau_i(X^0)\tau_j(X^0)\tau_k(X^0)|}\frac{h_k\epsilon_{ijk}}{\tau_k(X^0)},
\ee
where $\epsilon_{ijk}=1$ is totally antisymmetric and $\epsilon_{123}=1$. It was already found in \cite{AEM1}
that, in order to have homogeneous beta functions $\beta^g_{00}$ and $\beta^\phi$, one needs the following power
laws
\bea
\frac{d\phi(X^0)}{dX^0}&=&\frac{\phi_0}{X^0},\nn
\kappa(X^0)&=&\frac{\kappa_0}{(X^0)^2}.
\eea
From the first of the above equations, it becomes evident that the dilaton has a logarithmic dependence on the time $X^0$,
\begin{equation}
\phi = {\rm constant} + \phi_0\ln X^0
\label{logdil}
\end{equation}
In view of this, the antisymmetric tensor form (\ref{axion}) becomes 
\be\label{axiondilaton}
H_{\rho\mu\nu}(X^0) \propto e^{\frac{m}{\phi_0}\phi(X^0)} \varepsilon_{\rho\mu\nu\sigma}(X^0)\partial^\sigma b,
\ee
which shows a more familiar form, depending exponentially on the dilaton, 
appearing in standard string theory works \cite{ABEN,copeland}. In order to get homogeneous beta functions, the other
components of the metric also need to be power laws, and we write:
\be 
\tau_i(X^0)=\frac{-\kappa_0}{(X^0)^{n_i}}.
\ee
The antisymmetric three-form is then $H_{0ij}=-\kappa_0h_k(X^0)^{m-1+(n_k-n_i-n_j)/2}\epsilon_{ijk}$, and the
expressions for the one-loop beta functions are (no summation over space indices)
\bea
\beta_{00}^{g(1)}&=&\frac{-1}{4(X^0)^2} \sum_{i=1}^3 n_i^2 - \frac{1}{2}\sum_{i=1}^3 h_i^2 (X^0)^{2m+n_i-2} \nn
\beta_{ii}^{g(1)}&=&\frac{n_i}{4(X^0)^{n_i}}\left[4\phi_0+\sum_{i=1}^3 n_i \right]+\frac{1}{2} \sum_{j=1}^3 h_j^2
(X^0)^{2m+n_j-n_i} \nn \beta_{ij}^{g(1)}&=&-\frac{1}{2}h_ih_j(X^0)^{2m} \nn
\beta_{ij}^{B(1)}&=&\epsilon_{ijk}~h_k(X^0)^{-(\frac{n_i}{2}+\frac{n_j}{2}-\frac{n_k}{2}-m)}
\left[\frac{n_i}{4}+\frac{n_j}{4}-\frac{n_k}{4}-\frac{m}{2} + \phi_0\right]\nn
\beta^{\phi(1)}&=&-\frac{11}{3\alpha^{'}}+\frac{\phi_0}{4\kappa_0}\left[ 4\phi_0+\sum_{i=1}^3 n_i\right] -
\sum_{i=1}^3 \frac{h_i^2}{4 \kappa_0}(X^0)^{2m+n_i}
\eea
In order for these one-loop beta functions to be
homogeneous, it is necessary and sufficient to satisfy the constraints
\be\label{conditionhomogeneous}
2m+n_i=0~~~~\mbox{for those $i$ for which}~h_i\ne 0~,
\ee
Next we give the characteristic terms appearing at  second order in $\alpha^{'}$,
and check that the condition (\ref{conditionhomogeneous}) is respected.
\begin{itemize}

\item for $\beta^g_{00}$:
\bea \label{betasecond}
    R_{0\alpha\beta\gamma}R_0^{~\alpha\beta\gamma}&=&\frac{1}{16\kappa_0}(X^0)^{-2} \sum_{i=1}^3n_i^4\nn
    R^{\alpha\beta\rho\sigma}H_{0\alpha\beta}H_{0\rho\sigma}&=&-\frac{1}{4\kappa_0}\sum_{i,j,k=1}^3
    h_k^2n_in_j(X^0)^{2m-2+n_k}\nn H_{\rho\sigma0}H^{\sigma\alpha\beta}H^\rho_{~\beta\gamma}H^\gamma_{~\alpha
    0} &=&\frac{2}{\kappa_0}(X^0)^{4m-2}\sum_{i,j,k=1}^3h_i^2 (X^0)^{n_i}\sum_{l=1}^3 h_l^2(X^0)^{n_l}\nn
    \nabla_0H_{\alpha\beta\gamma}\nabla_0H^{\alpha\beta\gamma}&=&\frac{1}{2\kappa_0}
    \sum_{i,j,k=1}^3h_k^2\left(n_i+n_j-n_k-2m\right)^2(X^0)^{2m+n_k-2}\nonumber
\eea

\item for $\beta^g_{ii}$ (no summation on $i$): \bea
    R_{i\alpha\beta\gamma}R_i^{~\alpha\beta\gamma}=-\frac{1}{8\kappa_0}(X^0)^{-n_i}n_i^2\sum_{k=1}^3n_k^2\nonumber
    \eea

\item for $\beta^g_{ij}$, with $i\ne j$:
\bea
R^{\alpha\beta\rho\sigma}H_{i\alpha\beta}H_{j\rho\sigma}&=&
    \frac{n_k^2}{2\kappa_0}h_ih_j(X^0)^{2m}\nn H_{\rho\sigma
    i}H^{\sigma\alpha\beta}H^\rho_{~\beta\gamma}H^\gamma_{~\alpha j} &=&-\frac{8}{\kappa_0}(X^0)^{4m}
    h_ih_j\sum_{l=1}^3 h_l^2(X^0)^{n_l}\nonumber
\eea

\item for $\beta^B_{ij}$, with $i\ne j\ne k$:
\bea R_{i\gamma\alpha\beta}\nabla^\gamma
    H^{\alpha\beta}_{~~~j}&=& -\frac{n_i^2h_k}{8\kappa_0}(n_i+n_j-n_k-2m) (X^0)^{-(n_i+n_j-n_k-2m)/2}\nn
    \nabla_\gamma H_{\alpha\beta i}H_{j\rho}^{~~\alpha}H^{\beta\gamma\rho}&=&
    \frac{1}{2\kappa_0}(n_i+n_j-n_k-2m) \Big[h_k^3(X^0)^{(-n_i-n_j+3n_k+6m)/2}\nn
    &&+h_kh_i^2(X^0)^{(n_i-n_j+n_k+6m)/2}\Big]\nn &&+\frac{1}{2}(n_i-n_j+n_k-2m)
    h_kh_j^2(X^0)^{(-n_i+n_j+n_k+6m)/2}\nonumber
\eea

\item for $\beta^\phi$:
\bea \left(H_{\alpha\beta\gamma}H^{\alpha\beta\gamma}\right)^2&=&
    \frac{36}{\kappa_0^2}\sum_{i=1}^3h_i^4(X^0)^{4m+2n_i}\nn
    R_{\lambda\mu\nu\rho}R^{\lambda\mu\nu\rho}&=&\frac{1}{16\kappa_0^2} \sum_{i,j=1}^3 n_i^2(2n_i^2+n_j^2)\nn
    H_{\alpha\beta}^{~~~\mu}H^{\alpha\beta\nu}\nabla_\mu\nabla_\nu\phi&=&
    \frac{\phi_0}{\kappa_0^2}\left[\sum_{i,j=1}^3 n_ih_j^2(X^0)^{2m+n_j}\right ]\nn
    R^{\alpha\beta\rho\sigma}H_{\alpha\beta\lambda}H_{\rho\sigma}^{~~~\lambda}&=&
    -\frac{1}{2\kappa_0^2}\left[\sum_{i,j,k=1}^3 n_i(n_i+n_j)h_k^2(X^0)^{2m+n_k}\right ]\nn
    H_{\alpha\beta}^{~~~\mu}H^{\alpha\beta\nu}H_{\gamma\delta\mu}H^{\gamma\delta}_{~~~\nu}&=&
    \frac{4}{\kappa_0^2}(X^0)^{4m}\Bigg[\sum_{i,j=1}^3h_i^2h_j^2(X^0)^{n_i+n_j} \nn &~~~& +
    \sum_{i,j=1}^{3,\,i\neq j}h_i^2[3h_i^2(X^0)^{2n_i} + 4h_j^2(X^0)^{n_i+n_j}]\Bigg] \nn \nabla_\lambda
    H_{\alpha\beta\gamma}\nabla^\lambda H^{\alpha\beta\gamma}&=&\frac{3}{2\kappa_0^2}
    \sum_{i,j,k=1}^3h_k^2\left(n_i+n_j-n_k-2m\right)^2(X^0)^{2m+n_k}\nonumber
\eea

\end{itemize} As can be seen from the above expressions, the two-loop beta functions do not provide any
additional constraints over the ones found at one loop (\ref{conditionhomogeneous}). Also, higher orders of
the beta functions will not bring additional constraints, as can be seen by power counting: whatever power of
the Riemann or Ricci tensor we consider, multiplied by derivatives of the dilaton and/or the three-form field,
contracting the space-time indices with the metric or its inverse always leads to the expected power of $X^0$.
As explained in the introduction, conformal invariance is then satisfied for any non-vanishing value of the
amplitude $\phi_0$, and for any power $m$.

{\it Cosmological Properties.} The string frame metric $g_{\mu\nu}(X^0)$ is related to the Einstein frame
metric $g^E_{\mu\nu}(t)$ by \cite{ABEN}
\be
g^E_{\mu\nu}(t)dx^\mu dx^\nu=\exp\left\{-2\phi(x^0)\right\}g_{\mu\nu}(x^0)dx^\mu dx^\nu,
\ee
where $x^\mu$ denotes the zero mode of
$X^\mu$ and $t$ is the cosmic time, such that
\be\label{einsteinframe}
dt^2-\sum_{i=1}^3
a_i^2(t)(dx^i)^2=(x^0)^{-2\phi_0}\kappa_0 \left[\left(\frac{dx^0}{x^0}\right)^2-\sum_{i=1}^3
\frac{(dx^i)^2}{(x^0)^{n_i}}\right],
\ee
where $a_i(t)$ are the scale factors. Using the identity
(\ref{einsteinframe}), we find the following relation between the cosmic time $t$ and the coordinate $x^0$:
\be\label{tx0} 
t=\frac{\sqrt{\kappa_0}}{|\phi_0|}(x^0)^{-\phi_0}. 
\ee 
In the rest of the paper we concentrate, for concreteness, on
the case where $h_1=h_2=0$, $h_3=h\ne 0$ and take $n_1=n_2=n$. The configuration leading to homogeneous beta
functions is, in the string frame,
\bea\label{solution1}
g_{\mu\nu}(X^0)&=&\kappa_0~\mbox{diag}((X^0)^{-2},-(X^0)^{-n},-(X^0)^{-n},-(X^0)^{2m})\nn
H_{012}(X^0)&=&-\kappa_0
h(X^0)^{-1-n}\nn \phi(X^0)&=&\phi_0\ln(X^0).
\eea
From eqs.(\ref{einsteinframe}) and (\ref{tx0}), the scale
factors in the Einstein frame read then
\bea\label{scalefactors1} a_1(t)=a_2(t)&=&a_0~t^{1+n/(2\phi_0)}\nn
a_3(t)&=&\tilde a_0~t^{1-m/\phi_0}
\eea
where $a_0,\tilde a_0$ are constants, and the corresponding target space
is a power-law expanding Universe. Note that the case studied in \cite{AMT1} is more restrictive, since it was
done for $m=0$ only, which lead necessarily to a linearly expanding Universe along the $x^3$-direction .
In the present case we are able to discuss homogeneous and \emph{isotropic} solutions (\ref{scalefactors1}),
by choosing appropriate values for $n,m$ and $\phi_0$.

{\it Coupling to the Electromagnetic field.}
We couple the massless excitations of the string to a $U(1)$ gauge field by the introduction of the
following modified three-form field strength $\tilde H_{\rho\mu\nu}$
\cite{birefringence}: \be\label{atmod} \tilde H_{\rho\mu\nu}=H_{\rho\mu\nu}+\frac{1}{M}A_{[\rho}F_{\mu\nu]}, \ee
where $M$ has units of a mass scale. In our approach it can be equated~\cite{AMT1} to the string mass scale $M_s = (\alpha ')^{-1/2}$.
We concentrate here on the case
where $m=\phi_0=-1$ and $n=2$, leading to a four-dimensional Minkowski Universe. Indeed, 
on rescaling the space coordinates by $a_0,\tilde a_0$ and time by $\sqrt{\kappa_0}$, we obtain from
eqs.(\ref{tx0},\ref{solution1},\ref{scalefactors1}) the following configuration
\bea\label{config1}
g^E_{\mu\nu}&=&\eta_{\mu\nu} \nn H_{012}(t)&=&-\frac{\kappa_0 h}{t^3}\nn
\phi(t)&=&-\ln t.
\eea
The effective action in the Einstein frame is then \cite{low.ener.eff.}:
\bea\label{loweneffact}
S_{eff}&=&-\int d^4x
\sqrt{|g|}\left\lbrace \frac{44}{3\alpha^{'}}e^{2\phi}+
R-2\partial_\mu\phi\partial^\mu\phi\right.\nn
&&~~~~~~~~~~~~~~\left.-\frac{e^{-4\phi}}{12}\tilde H_{\rho\mu\nu}\tilde
H^{\rho\mu\nu} -\frac{e^{-2\phi}}{4}F_{\mu\nu}F^{\mu\nu}\right\rbrace,
\eea
where the coordinates $x^\mu$ are dimensionless, as they are rescaled by $\sqrt{\alpha^{'}}$. Note that the
first term in the effective action (\ref{loweneffact}), $-\frac{44}{3\alpha^{'}}e^{2\phi} =
\frac{2(D-26)}{3\alpha^{'}}e^{2\phi}$ is due to the non-critical dimension and from a low-energy observer represents 
a target-space vacuum contribution. Notice that for sub-critical-dimension strings (as is the case of our four-dimensional 
configuration), the dark energy contribution is negative (anti-de-Sitter) type, relative to the Einstein curvature term. 
This could imply troubles for the phenomenology of the model, unless this negative contribution to the vacuum energy is 
compensated by other terms in the effective action. This issue is not resolved in the current paper, although it should 
be noted that the time-dependent dilaton implies that this vacuum energy decays as $\exp(2\phi)=1/t^2$.

The presence of a time-dependent dilaton implies a
a cosmic-time-dependent fine structure constant \cite{damour}, and the phenomenology of the latter was already discussed in \cite{AMT1}. 
Indeed, plugging the
configuration (\ref{config1}) in the low energy effective action (\ref{loweneffact}) we find that the relevant
effective action for the gauge field is
\be\label{gauge1} 
S_{gauge}=\int dt d\vec x \left\lbrace
-\frac{t^2}{4}F_{\mu\nu}F^{\mu\nu} +\epsilon t A_{[0}F_{12]}\right\rbrace,
\ee
where $\epsilon=\kappa_0h/M_s$ and terms of order $1/M_s^2$ have been ignored.

The equations of motion for the electromagnetic field in the background
(\ref{config1}) read then
\bea\label{newMax1}
\vec\nabla\cdot\vec B&=&0\\
\vec\nabla\cdot\vec E&=&\frac{2\epsilon}{t}B_3\nn
\vec\nabla\times\vec E&=&-\partial_t\vec B\nn
\vec\nabla\times\vec B&=&\partial_t\vec E+\frac{2}{t}\vec E +\frac{\epsilon}{t}\vec E_\bot-\frac{\epsilon}{t^2}\vec A_\bot,\nonumber
\eea
where $\vec
E_\bot=(-E_2,E_1,0)$ and $\vec A_\bot=(-A_2,A_1,0)$. Note that the gauge dependence of these equations is due to
the gauge fixing in eq.(\ref{atmod}), and is compensated, during a gauge transformation in the original space time action
(\ref{loweneffact}), by a simultaneous transformation of the Kalb-Ramond field.

We are looking for a plane wave solution of eqs.(\ref{newMax1}), propagating
in the $x^3$-direction, thus depending on $t$ and $x^3$. Ignoring the terms proportional to $1/t^2$, and using the complex notation
${\bf E}=E_1+iE_2$ in the plane $(x^1,x^2)$, we obtain the following equation of motion
\be\label{evoleq}
\ddot{\bf E}-{\bf E}^{''}+\frac{2+i\epsilon}{t}\dot{\bf E}=0,
\ee
where a dot represents a time
derivative and a prime represents a space derivative. It can be easily seen that the ansatz
${\bf E}={\bf E_0}t^a\cos(t-x^3)$, where ${\bf E_0}$ and $a$ are complex constants, satisfies equation (\ref{evoleq}) for
$a=-1-i\varepsilon/2$, up to terms of higher order in $1/t$, which are neglected in our approach. In this
approximation, the solution is then
\be\label{solE}
{\bf E}=\frac{{\bf E_0}}{t}\exp\left( -i\frac{\epsilon}{2}\ln t\right)\cos(t-x^3).
\ee
Due to the extra time-dependence induced by the dilaton and axion (compared to the usual plane wave solution of Maxwell's equations),
the magnetic field ${\bf B}$ is in principal {\it not} perpendicular to the electric field. But the component of ${\bf B}$ in the
direction of ${\bf E}$ is of higher order in $1/t$, which is neglected in our approximation. Indeed, from the modified
Maxwell's equations (\ref{newMax1}), we obtain
\bea
{\bf B}&=&i{\bf E_0}\exp\left( -i\frac{\epsilon}{2}\ln t\right)\left(-\frac{\cos(t-x^3)}{t}+\frac{\sin(t-x^3)}{t^2}\right) \nn
&&-\frac{\varepsilon}{2t^2}{\bf E_0}\exp\left( -i\frac{\epsilon}{2}\ln t\right)\sin(t-x^3)\nn
&=&-i{\bf E}+{\cal O}(1/t^2),
\eea
such that, in our approximation, electric and magnetic fields are perpendicular.

One can identify the roles of the dilaton and the antisymmetric tensor in the solution (\ref{solE}): the dilaton leads
to an amplitude decreasing as $1/t$, while the Kalb-Ramond field
implies a polarization angle varying logarithmically with the cosmic time:
\be\label{oa}
\Delta (t) =|\mbox{arg}({\bf E})-\mbox{arg}({\bf E^\star})|=\epsilon\ln t, \qquad \epsilon=\kappa_0h/M_s,
\ee
where $t$ can be seen as the time interval between emission and observation of the electromagnetic waves.
Notice that the antisymmetric tensor field is responsible for the optical activity $\Delta(t) $ even in a Minkowski background,
as is our case here. We notice that the result (\ref{oa}) is identical to that for the optical activity in the cosmically anisotropic situation considered in \cite{AMT1}, where the $x^3$-direction of the Universe was expanding linearly in time.
Because of the Minkowski situation, the time interval must be assumed sufficiently small in a cosmic scale, so as
global curvature effects are ignored when one is discussing the relevant phenomenology in order to constrain the
characteristic parameter $\epsilon $.

In this respect, it is interesting to compare our results on the optical activity (\ref{oa}), stemming from the
non-perturbative in $\alpha ' $ four-dimensional background configuration (\ref{solution1}), with higher-dimensional
situations in other string- or brane- inspired scenarios, such as the Randall-Sundrum model \cite{birefringence}.
There, the  case of axion-induced  optical activity in (compactified) four-dimensional
Minkowski brane worlds was considered, among other examples.
For relatively small time scales, the optical activity was found  to be proportional to the comsic time interval
from emission till observation of the wave. In contrast, in our case this dependence is logarithmic, as a consequence of the
power-law configurations we need, in order to satisfy conformal invariance in a non-perturbative way.

In realistic situations, the cosmic curvature has to be taken into account by considering the case of a proper isotropic
Robertson-Walker (RW) background of an expanding Universe, with dark energy contributions. However, formally, to incorporate
such backgrounds in our non-perturbative framework is not a trivial task. So far \cite{AEM1} we have only derived power-law
expanding Universes, with non trivial dilatons.
Such Universes are characterized (perturbatively in $\alpha ' $) by a relaxing to zero dark energy, due to the non-trivial
dilaton potential terms (arising from the $D-26$ non-critical contributions to the target-space effective action and/or from
string loops~\cite{ABEN,west,veneziano}. However, the $\alpha '$-non-perturbative situation within our framework is still to
be worked out. One, of course, may do preliminary phenomenological studies of the antisymmetric-tensor-induced optical effects
by assuming that the effects of the antisymmetric-tensor field on the cosmic background are negligible, as was done in
\cite{birefringence}, and discuss the optical activity in such scenarios. However, within our $\alpha '$-non-perturbative
homogeneous in time situation (\ref{solution1}), such an assumption is most likely invalid, and one needs to understand
the emergence of RW cosmology in the presence of antisymmetric tensors and dilaton fields in a self consistent way.
This is in progress.

\section*{Acknowledgements} 

The work of N.E.M. is partially supported by the European Union through the FP6 Marie Curie Research and Training Network \emph{UniverseNet} (MRTN-CT-2006-035863).


\begin{thebibliography}{99}


\bibitem{AEM1}
  J.~Alexandre, J.~Ellis and N.~E.~Mavromatos,
  JHEP {\bf 0612} (2006) 071
  [arXiv:hep-th/0610072].

\bibitem{metsaev} R.~R.~Metsaev and A.~A.~Tseytlin,
  Nucl.\ Phys.\ B {\bf 293}, 385 (1987).



\bibitem{AMT1} J.~Alexandre, N.~E.~Mavromatos and D.~Tanner,
  New J.\ Phys.\  {\bf 10}, 033033 (2008)
  [arXiv:0708.1154 [hep-th]].




\bibitem{copeland}
  E.~J.~Copeland, A.~Lahiri and D.~Wands,
  Phys.\ Rev.\  D {\bf 50} (1994) 4868
  [arXiv:hep-th/9406216];
  Phys.\ Rev.\  D {\bf 51} (1995) 1569
  [arXiv:hep-th/9410136].



\bibitem{nodland}
  B.~Nodland and J.~P.~Ralston,
  Phys.\ Rev.\ Lett.\  {\bf 78} (1997) 3043
  [arXiv:astro-ph/9704196].





\bibitem{ABEN}
  I.~Antoniadis, C.~Bachas, J.~R.~Ellis and D.~V.~Nanopoulos,
  Phys.\ Lett.\  B {\bf 211} (1988) 393;
  I.~Antoniadis, C.~Bachas, J.~R.~Ellis and D.~V.~Nanopoulos,
  Nucl.\ Phys.\  B {\bf 328} (1989) 117.





\bibitem{birefringence}
  P.~Das, P.~Jain and S.~Mukherji,
  Int.\ J.\ Mod.\ Phys.\  A {\bf 16} (2001) 4011
  [arXiv:hep-ph/0011279];
  S.~Kar, P.~Majumdar, S.~SenGupta and A.~Sinha,
  Eur.\ Phys.\ J.\  C {\bf 23} (2002) 357
  [arXiv:gr-qc/0006097];
  D.~Maity, S.~SenGupta and S.~Sur,
  Phys.\ Rev.\  D {\bf 72} (2005) 066012
  [arXiv:hep-th/0507210].




\bibitem{low.ener.eff.}
  E.~S.~Fradkin and A.~A.~Tseytlin,
  Phys.\ Lett.\  B {\bf 158} (1985) 316,
C.~G.~.~Callan, E.~J.~Martinec, M.~J.~Perry and D.~Friedan,
  Nucl.\ Phys.\  B {\bf 262} (1985) 593.

\bibitem{damour} T.~Damour and A.~M.~Polyakov,
  Nucl.\ Phys.\  B {\bf 423}, 532 (1994)
  [arXiv:hep-th/9401069];
  Gen.\ Rel.\ Grav.\  {\bf 26}, 1171 (1994)
  [arXiv:gr-qc/9411069].




\bibitem{west} J.~R.~Ellis, N.~E.~Mavromatos, D.~V.~Nanopoulos and M.~Westmuckett,
  Int.\ J.\ Mod.\ Phys.\  A {\bf 21}, 1379 (2006)
  [arXiv:gr-qc/0508105].


\bibitem{veneziano} M.~Gasperini and G.~Veneziano,
  Phys.\ Rept.\  {\bf 373}, 1 (2003)
  [arXiv:hep-th/0207130];
  M.~Gasperini, F.~Piazza and G.~Veneziano,
  Phys.\ Rev.\  D {\bf 65}, 023508 (2002)
  [arXiv:gr-qc/0108016].

\end{thebibliography}
\end{document}